\documentclass[copyright,creativecommons]{eptcs}

\usepackage{amssymb}
\usepackage{amsmath}

\usepackage[dvips]{graphicx}

\providecommand{\event}{}
\providecommand{\volume}{2}
\providecommand{\anno}{2009}
\providecommand{\firstpage}{3}
\usepackage{breakurl}

\title{A Peer to Peer Protocol for Online Dispute Resolution over Storage Consumption}
\author{Ahmed Mihoob \qquad Carlos Molina-Jimenez
\institute{School of Computing Science\\
 Newcastle University, UK}
\email{\{a.m.mihoob,carlos.molina\}@ncl.ac.uk}
}

\def\titlerunning{A Peer to Peer Protocol for Online Dispute Resolution over Storage Consumption}
\def\authorrunning{A.~Mihoob \& C.~Molina-Jimenez}

\begin{document}
\maketitle

\begin{abstract}
In bilateral accounting of resource consumption both the consumer 
and provider independently measure the amount of resources consumed 
by the consumer. The problem here is that potential disparities 
between the provider's and consumer's accountings, might lead to 
conflicts between the two parties that need to be resolved. We argue 
that with the proper mechanisms available, most of these conflicts can 
be solved online, as opposite to in court resolution; the design of 
such mechanisms is still a research topic; to help cover the gap, in 
this paper we propose a peer--to--peer protocol
for online dispute resolution over storage consumption. 
The protocol is peer--to--peer and takes into 
consideration the possible causes (e.g, transmission delays, 
unsynchronized metric collectors, etc.) of the disparity between the 
provider's and consumer's accountings to make, if possible, the two 
results converge.
\end{abstract}

\section{Introduction}
The general scenario of our research interest is the consumption
of computing resources (storage, bandwidth, computation, etc.) offered
by providers to remote users (consumers) over the Internet. 
The consumer regards the
resources as a service reachable through a user's interface 
and pays for it on a pay--per--use basis. Central to this scenario
is resource consumption accounting. Currently, most providers
use unilateral provider--side accounting based on metrics 
collected by devices deployed within the providers' premises.
An alternative and innovative approach is bilateral accounting 
where both the consumer and provider independently measure resource 
consumption and verify the parity of the accounting 
results~\cite{MolinaESBE2008}. A
potential problem here is
the emergence of potential conflicts derived from divergences 
between the independently produced accounting results. 
The practicality
of bilateral accounting depends on whether most conflicts can be
solved online, as opposite to off-line resolution; this issue is
still an open research question. To help cover the gap, in this
paper we present an online peer--to--peer protocol for
dispute resolution over resource consumption. To meet space
and time constraints and focus the discussion on an specific 
and practical example, we deal only with storage consumption
and in particular we concentrate on a rather simple scenario 
where the consumer can only upload data to an incremental
storage service. 

\begin{figure}[h]
        \centering
                \includegraphics[width=0.80\textwidth]{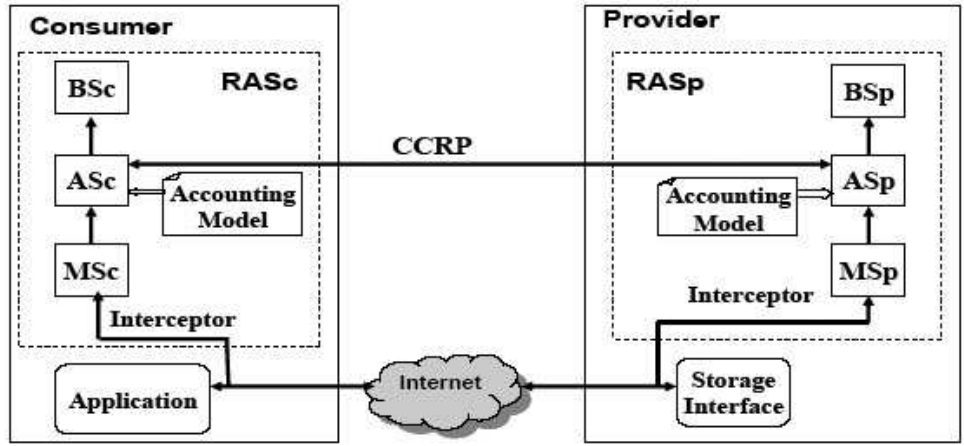}
        \caption{A bilateral storage accounting system.}
        \label{fig:scenario}
\end{figure}

An abstract view of our scenario of study is shown in
Figure~\ref{fig:scenario} where \emph{Provider} represents
an storage service and \emph{Consumer} represents the
consumer of the service which can be a single individual or
a large enterprise with scores of employees or a university with 
thousands of students. As shown in the figure, consumer and provider 
deploy their own resource accounting services 
($RAS_c$ and $RAS_p$, respectively) within theirs respective
infrastructures. A RAS is composed of three components: a 
metering service (MS) responsible for collecting raw metering data
about storage consumption; an accounting service (AS) that retrieves 
the metering data and applies an accounting model to produce 
accounting data; and a billing service (BS) that on the
basis of the accounting data provided by the AS and
pricing policies (e.g., discounts to golden customers,
fines to late payments, etc.) produces
the actual bill, say monthly, for the consumer.  
As shown in the figure, the consumer can access the service
only through a storage interface---the service interface
offered to consumers. 

The CCRP (Comparison and Conflict Resolution Protocol) is
the central topic of this paper and represents the protocol
that the consumer and provider execute when conflicts over
storage consumption emerge, with the intention of solving 
them online. As shown in the figure, the protocol is executed
by the accounting services when the difference between their
independently produced accounting results is greater than an
agreed upon value. 
We anticipate several sources of conflicts. For example,
a primary source of conflics is the accounting model
used by consumer and provider to compute accounting data. 
As some in the figure, to avoid this problem, in this paper
we require that both consumer and provider use the
same accounting model which is publish by the provider.
Such a model is basically an algorithm that aggregates
raw metering data (e.g., 300000 upload this week) and 
converts it into accounting records over agreed upon consumption 
intervals (10 Mbytes/Mon, 12Mbytes/Tue, etc.).    
Another source of potential conflicts is the
techniques used by consumer and provider to collect 
metering data with their MSs. For example, the consumer
might rely on interceptors whereas the provider---with
unrestricted access to its infrastructure---might 
measure storage consumption directly from its 
file servers. At this stage of our
research, and as shown in the figure, we assume that
both consumer and provider use interceptors to
collect metering data. There are others sources of
potential conflicts, yet we will concentrate on
the transmission time of requests and
the accuracy of accounting intervals as these two
parameters are the most relevant to the scenario
of our interest.  We will discuss the CCRP
(as well as the interceptors) at large in the following
sections.

\section{The Protocol}
The CCRP is an peer--to--peer conflict resolution
protocol in the sense that it is executed between
the two conflicting parties without the intervention
of a third one, such as a referee or arbitrator. It
is an online protocol in that it is executed 
immediately upon the detection of a conflict and 
as the service is delivered to the consumer. More
importantly, it is an evidence--based protocol in the
sense that, on the basis of evidence provided by
the conflicting parties, it tries to identify the
source of the divergency between the two accounting
results. It is worth mentioning that this approach
departs from conflict resolution protocols
based on Utility Theory which are interest--based in
that, they take into consideration the
conflicting parties' preferences and 
tradeoffs~\cite{ZCXu2008}.

\section{Assumptions}
We admit that the CCRP is still under development.
In this paper we explore its feasibility in a
very simple incremental storage consumption scenario described by the assumption
discussed below. As explained in Section~\ref{FutureWork},
we are planning to relax these assumptions in the future
to generalise our scenario. 
We believe that the fundamental ideas (e.g.~the architecture shown in
Figure~\ref{fig:Architecture}) discussed in the 
current scenario will still hold in a more general one.

The scenario of study can be described by the following assumptions:

\begin{enumerate}
\item The provider offers an incremental storage service where
      \emph{upload file} is the only operation available to
      the consumer to alter its storage space and each
      execution results in the creation of a new file. 
      This service is far from being a general scenario, yet it
      is still of some practical interest (e.g.~in archival 
      storage~\cite{Storer2008}) and
      more importantly, it is good enough to explain our
      ideas.
\item All the consumer's upload operations are requested from
      within its premises (see Figure~\ref{fig:scenario}); consumers
      with laptops that roam outside the premises are not considered.
\item The service is delivery continuously over an agreed period
      of time, for example, over a year.
\item In the interest of accountability, the total period is 
      divided into Consumption Intervals (CI) with 
      SP and EP (Start Point and End Point, respectively)
      determined by the provider.
\item Zero o more requests can issued by the consumer during
      the duration of each consumption interval. 
\item There are no gaps in the accounting line. Except for the
      last interval, the end of a given interval correspond to 
      the start of the next one.  
\item The consumer and provider independently produce their
      accounting record about the storage consumed over
      each consumption interval. The two independently produced
      records do not necessarily match.
\item The CCRP is executed for each consumption interval to
      compare the two independently produced accounting
      records and to try to solve potential conflicts. 
\item The provider's and consumer's clocks are synchronized.
\item The interceptors used by the consumer and provider
      are deployed as shown in Figure~\ref{fig:scenario} to
      intercept each consumer's request.
\item The $MS_{c}$ and $MS_{p}$ collect the following data
      about each request: \emph{Request id}, \emph{Request Time Stamp (RTS)}
      and \emph{Bytes Transferred per Request (BT)}. In addition, 
      $MS_{p}$ also collects \emph{Request Received Time (RRT)}.
\end{enumerate}

\section{Accounting Model}
\label{AccountingModel}
The accounting model is used by both the consumer and
provider to calculate the storage consumed  
by each request issued by the consumer. Under the assumption
that the provider relies on conventional file systems
to implement his service, our accounting model 
considers the number of bytes uploaded by the request
and the configuration parameters of the provider's
file system.

The number of bytes transferred by each request ($BTreq_i$) is
determined by the interceptors after intercepting and
examining the request.

The configuration parameters are inherent to the 
file system. In Our accounting model we consider
the amount of metadata ($MD$) associated to each file
and the size of the disk chunk ($ChSize$)---also called, size of
disk cluster. Typical values of $MD$ and  $ChSize$ are, 
respectively, 2KB and 4KB. In this order,
the number of chunks consumed by a request 
can be calculated by equation~\ref{Eq:NofCh}.

\begin{equation} \label{Eq:NofCh} 
NofCh_{i}=\frac{BTReq_{i}+MD}{ChSize} 
\end{equation} 

It follows that the storage consumed by a given request can be
calculated by equation~\ref{SCUF}:

\begin{equation} \label{SCUF}
  SCUF= RoundUp(NofCh) * ChSize 
\end{equation}

\emph{RoundUp} represents a round up operation to
the nearest integer and counts for the fact that 
disk chunks are allocated only in whole units.

\begin{figure}[h]
	\centering
		\includegraphics[width=0.90\textwidth]{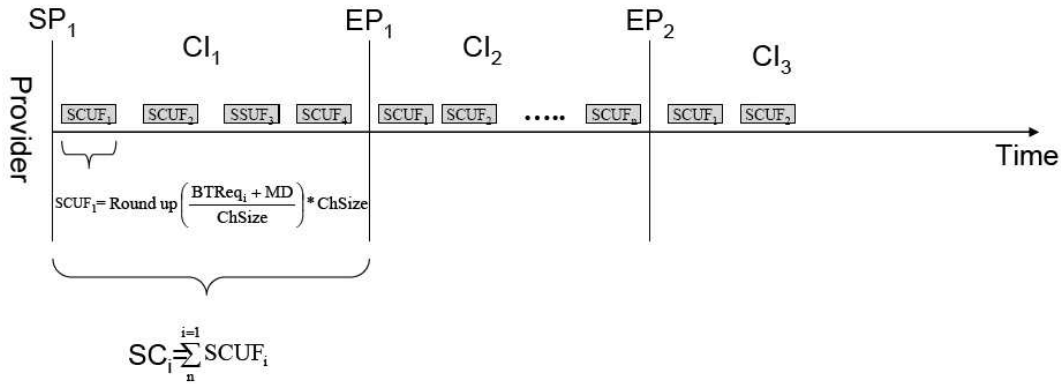}
	\caption{Storage consumed within each consumption interval.}
	\label{fig:ConsumptionIntervals}
\end{figure}

The amount of storage consumed within each consumption interval
can be calculated as the sum of the storage consumed by
each request issued within the interval; we represent
it by equation~\ref{SC} and show it graphically in 
Figure~\ref{fig:ConsumptionIntervals}.

\begin{equation} \label{SC}
 SC= \sum_{i=1}^{n} SCUF_{i}
\end{equation}

\section{Two Potential Causes of Disparities over Storage Consumption}
\label{CausesOfConflicts}
In this section we will explain how mismatches between 
the consumer's and provider's accounting intervals can
results in disparities between the consumer's an
provider's accounting results. Likewise, we will discuss
how transmission time impacts the accounting results
produced by the two parties. 

\subsection{Consumption Interval}
\begin{figure}[h]
	\centering
		\includegraphics[width=0.45\textwidth]{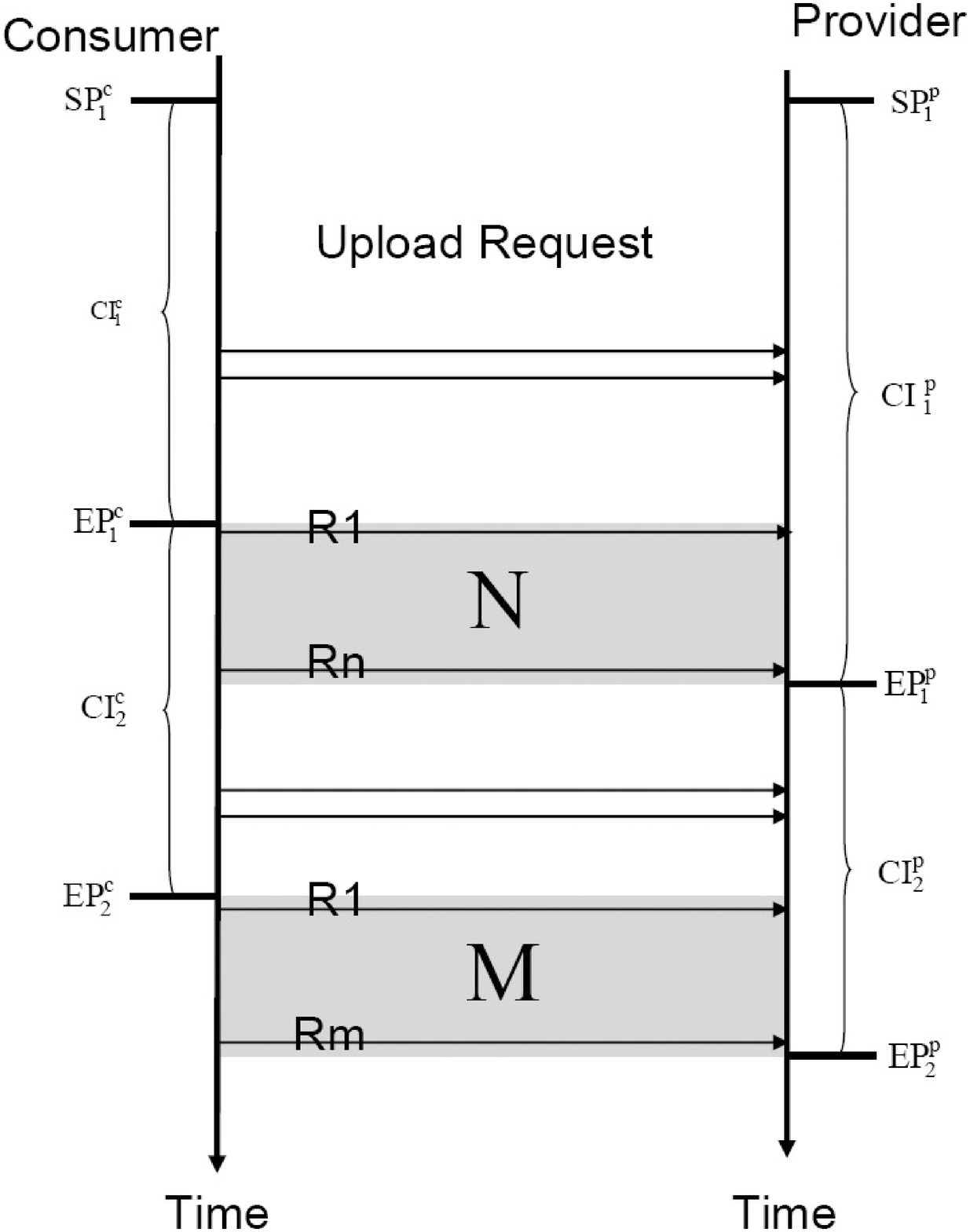}
	\caption{Mismatch between consumer's and provider's consumption intervals.}
	\label{fig:ImpactConsumptionIntervals}
\end{figure}

The impact of potential mismatches between the consumer's and
provider's consumptions intervals is shown in 
Figure~\ref{fig:ImpactConsumptionIntervals}. In the
figure, $CI$ and $R$ stand, respectively, for
 for Consumption Interval and Requests. Similarly,
$SP$ and $EP$ stand, respectively, for Start Point and End 
Point of a given interval. We refer with superscripts
$c$ and $p$, to consumer and provider, respectively.
Subscripts represent the sequence number of the interval; for
example, $SP_{1}^{c}$ represents the start point of 
the consumer's interval number one. Notice that for
simplicity, the figure assumes that the transmission 
time of the requests is zero.

As suggested by the figure, it is quite possible
that for a given interval the consumer's 
Start Point (SP) and End Point (EP) do not match 
the provider's. Such a mismatch is very likely to
result in divergencies between the consumer's and
provider's accounting records for the interval
under question. In the figure, $EP_{1}^{p} > EP_{1}^{c}$,
consequently, $CI_{1}^{p}$ includes $N$ requests
more than the consumer's  $CI_{1}^{c}$. Naturally,
the length of the divergency between the two
results depends on the amount of bytes transferred
in each requests and more importantly, on the
value of $N \geq 0$. As shown in the figure,
length of the divergency for $CI_2$ depends
on the value of $N$ and $M \geq 0 $.
 
In practice, this situation can arise when
the provider does not offer precise information
to the consumer about when to start and end a 
given consumption interval.  
For instance, most storage providers, like Amazon~\cite{Amazon}, 
and Nirvanix~\cite{Nirvanix} do not offer (e.g.~in their Service 
Level Agreements) sound accounting models to their 
customers. For example, in Amazon S3 service, the storage
consumed by a given customer is calculated as 
follows:  Amazon checks at least twice a day  
the consumer's storage space, it measures the amount 
of storage occupied by a consumer's buckets and 
multiplies the result by the amount of time elapsed since the 
last check. However, Amazon S3 does not state 
exactly when (in time units) they undertake their measurement. 

As discussed in Section~\ref{AccountingModel}, the
consumer and provider calculate storage consumption 
within a given consumption interval by 
equation~\ref{SC}. The requirement to make the
two independently produced results converge is that
both parties use exactly the same SP and EP for the
interval under question. We anticipate two possible
solutions to this problem. An alternative is to
keep the consumer's and provider's metering services
strictly synchronised; for example, the provider can
notify the consumer when each consumption interval
starts and ends. As shown in the pseudocode
presented in Section~\ref{Protocol}, in this paper we 
explore a second alternative where the parties
exchange their SP and EP upon detection of
conflicts between their accounting results.

\subsection{Transmission Time}
\begin{figure}[h]
	\centering
		\includegraphics[width=0.60\textwidth]{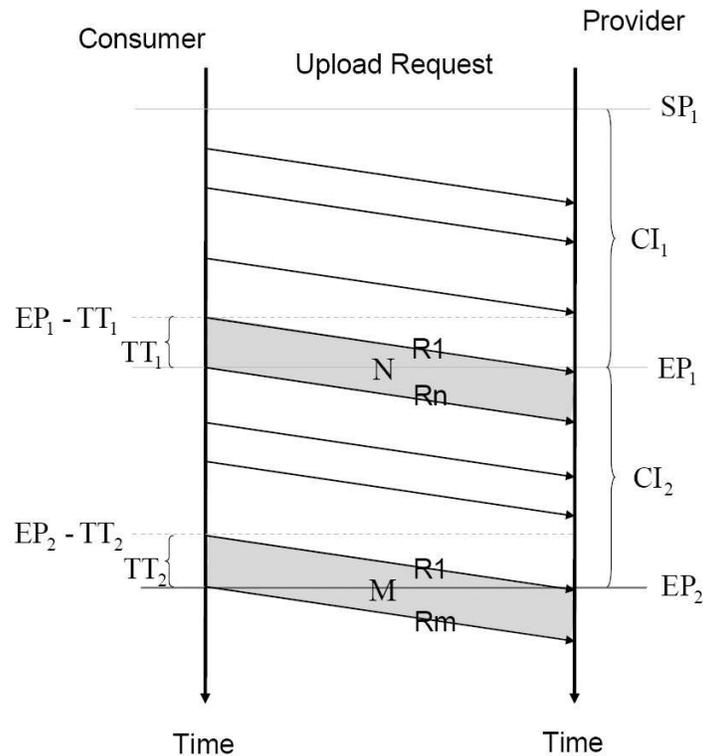}
	\caption{Impact of transmission time on storage accounting.}
	\label{fig:TransmissionTime}
\end{figure}

We define transmission time (TT) as the time it takes
a consumer's request to travel from the consumer to
the provider. In practical applications, TT is normally
greater than zero, say of the order of 100 milliseconds.
In Figure~\ref{fig:TransmissionTime}, TT represents
the average transmission time. As shown graphically,
this parameter can cause divergencies between the
consumer's and provider's accounting results for
a given consumption interval. For the sake of simplicity,
let us assume that the consumer's and provider's 
SP and EP of a given interval are synchronised. 
Under this assumption, convergency between the
consumer's and provider's accounting records can
be achieved by compensating the provider's results
by the amount of memory consumed by the requests in the wire, that is, 
requests issued in a given interval but received and counted in
the following due to TT.

Let us take an arbitray interval $CI_i$. 
The consumer can calculate its storage
consumption by equation~\ref{SC}. However,
to compensate for $TT$, the provider would 
need to use equation~\ref{eq:TT}.

\begin{equation} \label{eq:TT}
 SC= \sum_{i=1}^{n} SCUF_{i} + \mid N-M \mid
\end{equation}

where $N$ is the amount of storage consumed by requests
issued and counted by the consumer in interval $CI_{i-1}$  but received
and counted by the provider in interval $CI_i$ due to 
the effect of TT, in the figure this time gap is shown as
$TT_1$. Similarly, $M$ 
is the amount of storage consumed by requets issued and
counted by the consumer in interval $CI_i$ but to be 
received and counted by the provider in interval $CI_{i+1}$,
due to TT; in the figure, this time gap is shown as
$TT_2$. Both $N$ and $M$ can be calculated by
equation~\ref{SC}. Notice that for the first interval
$N$ is to be taken as $N=0$. 

An equivalent alternative to compensate the provider's
accounting results is for the provider (or the consumer)
to shift its consumption interval to count for $TT$. 
This is the strategy taken in the pseudocode presented
in Section~\ref{Protocol}. 

\vspace{0.2812in}
It is worth keeping in mind that accounting records
can be impacted simultaneously by both asynchrony
of consumption interval and transmission time. The
protocol presented in Section~\ref{Protocol} handles
the two potential sources of conflicts separately. First
it tries to match the results by considering the 
asynchrony of the interval; if it fails, it takes 
TT into account. Failure to produce matching
results leads to offline dispute resolution.

\section{The Model}
A crucial problem in accounting of storage consumption is the 
generation of non-repudiable evidence about
the consumption. We address this issue with the help of 
the piece of middleware for Non-Repudiable (NR) information 
sharing presented in~\cite{Cook2002,Cook2006}. The fundamental
idea is that the middleware provides multi-party, 
non--repudiable agreement to updates to shared 
information which can be maintained in a 
distributed manner with each party holding a copy. 
Essentially, one party proposes a new value for the 
state of some information and the other parties sharing 
the information subject the proposed
value to application--specific validation. If all 
parties agree to the value, then the
shared view of the information is updated accordingly. 
Otherwise, the shared view of the information remains 
in the state prior to proposal of the new value. 

\begin{figure}[h]
	\centering
		\includegraphics[width=0.70\textwidth]{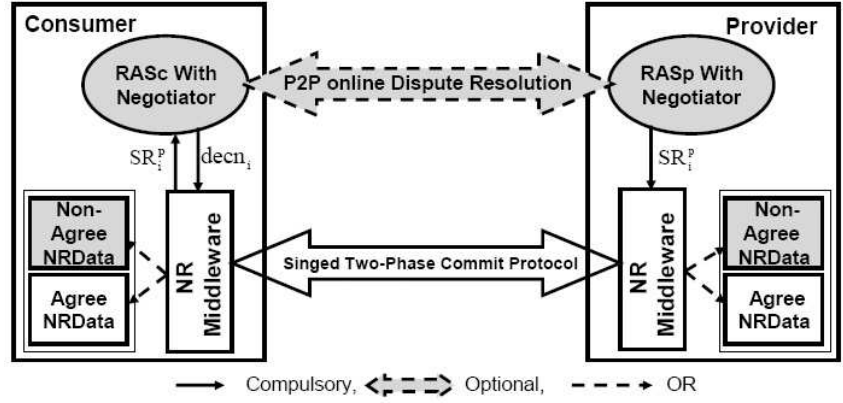}
	\caption{Architecture to support the online dispute resolution protocol.}
	\label{fig:Architecture}
\end{figure}

The architecture of our solution is shown in 
Figure~\ref{fig:Architecture}.
NR Midleware represents the non--repudiable
middleware. Similarly, $RAS_c$ and $RAS_p$ represent, respectively,
the consumer's and provider's resource accounting systems.
Non-Agreed NRData and Agreed NRData are
files to store, respectively, non-agreed and agreed accounting 
records, as determined by the P2P online Dispute Resolution 
protocol. Records from the Non-Agreed log can be used
in case of offline dispute resolution.

The signed two-phase commit protocols works as
follows:

\begin{enumerate}
\item  The provider $RAS_p$ calculates the 
accounting record  $SR_{i}^{p}$ for a given
consumption interval $CI_i$ and sends it to
its NR Midleware  
which produces non--repudiation of its 
origin $NRO(SR_{i}^{p})$ and sends 
$NRO(SR_{i}^{p})$ and $SR_{i}^{p}$ to the consumer. 

\item  The consumer's NR middleware validates $SR_{i}^{p}$ 
and $NRO(SR_{i}^{p})$ and sends $SR_{i}^{p}$ 
to its $RAS_c$.

\item  $RAS_c$ produces an accounting record $SR_{i}^{c}$
for $CI_i$, compares it with the $SR_{i}^{p}$ and 
produces a decision, $decni$. $decni$ is essentially a 
binary \emph{Yes} or \emph{No} value. If $decni=Yes$, 
$RAS_c$ sends $decni$ to the consumer NR middleware, 
otherwise, it triggers the online P2P dispute 
resolution. When this protocol is completed
$decni$ is sent to the consumer's NR middleware. 

\item  The consumer's NR middleware sends the decision 
$decni$, non-repudiation of receipt of $SR_{i}^{p}$, 
$NRR(SR_{i}^{p})$ and non--repudiation of origin 
of the decision $NRO(decni)$. 

\item  The provider NR middleware validates $decni$, 
$NRO(decni)$ and $NRR(SR_{i}^{p})$. The protocol 
terminates with the provider sending non--repudiation 
of receipt of the validation decision to the consumer 
$NRR(decni)$. 
\end{enumerate}

\subsection{The Provider's Resource Accounting Service}
\begin{figure}[h]
	\centering
		\includegraphics[width=0.70\textwidth]{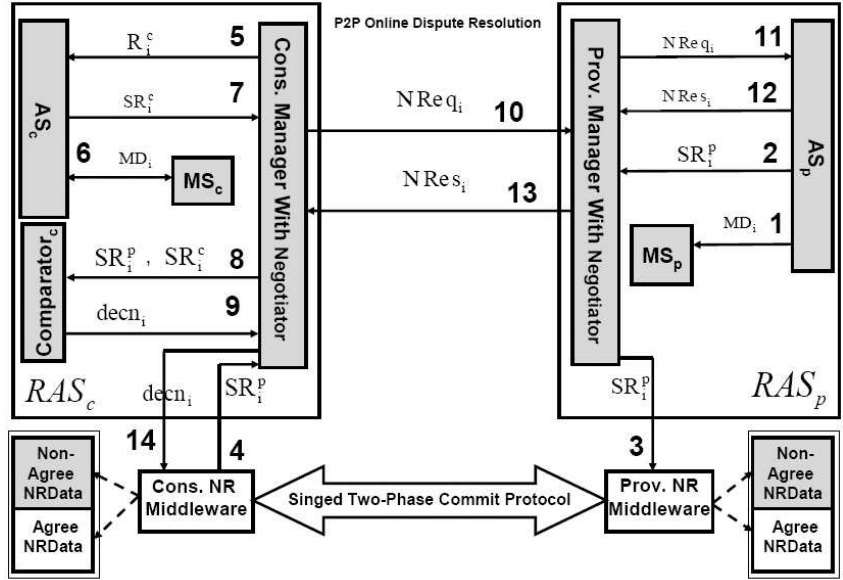}
	\caption{Architecture and online dispute resolution protocol.}
	\label{fig:Architecture2}
\end{figure}

We will elaborate on how the provider's resource accounting
service work, with the help of Figure~\ref{fig:Architecture2}
which is an expansion of Figure~\ref{fig:Architecture}.
$RAS_p$ is a service with a negotiator used by the provider to collect 
data, compute and negotiate resource consumption. $RAS_p$ 
consists of a Manager With Negotiator ($MWN_p$), 
Metering Service ($MS_p$) and Accounting Service ($AS_p$). 
The $MS_p$ collects data about resource consumption 
caused by all uploaded requests and stored them in 
a permanent file. It records the following data 
\emph{user Id, request Id, request time stamp, 
request arrived time and number of bytes transferred per request}. 

$AS_p$ determines the (SP, EP) for each $CI_i$, obtains the 
metered data from $MS_p$, calculates 
the average $TT$ and produces a standard 
accounting record $SR_{i}^{p}$ for each $CI_{i}^{p}$.
The $AS_p$ uses equation~\ref{eq:TTreq} to compute TT of
a request:

\begin{equation} \label{eq:TTreq}
 TT= RequestArrivalTime - RequestTimeStamp
\end{equation}

The average TT is calculated by equation~\ref{eq:TTave}.

\begin{equation} \label{eq:TTave}
TTave= \left(\sum _{i=1}^{n}TT_{i} \right) * \frac{1}{n} 
\end{equation}

The $MWN_p$ obtains $SR_{i}^{p}$ from its $AS_p$ and sends 
it to the provider NR middleware. Another responsibility of  
of $MWN_p$ is to negotiate with the consumer $RAS_c$. In 
the negotiation steps, the $MWN_p$ receives negotiation 
requests with the consumer's accounting 
parameters and the negotiator counter from the $RAS_c$. $MWN_p$ 
obtains the negotiator requests and sends it to 
$AS_p$ who compares them with its accounting parameters. 
The $AS_p$ obtains the parameter or parameters which 
cause the conflict, updates the negotiator counter 
and sends the negotiation response to the $MWN_p$ who 
sends it to the $RAS_c$ and waits for new negotiation requests.

\subsection {Consumer Resource Accounting Service}
Similarly to the provider's resource accounting
system, the $RAS_c$ is a service with a
negotiator that is used by the consumer to collect 
data, compute, negotiate and produce decisions about 
storage consumption for each CI. $RAS_c$  consists of 
a Manager With Negotiator ($MWN_c$), Metering Service 
($MS_c$), Accounting Service ($AS_c$) and a 
$Comparator_c$ (see Figure~\ref{fig:Architecture2}). 
The $MS_c$ collects data about resource consumption 
caused by all uploaded requests and stores them in 
a permanent file. It records the following data 
\emph{request Id, request time stamp, number of 
bytes transferred per request}. $RAS_c$ works as follows:

\begin{enumerate}
\item  $MWN_c$ receives the provider's standard 
accounting record $SR_{i}^{p}$ from the consumer NR middleware.

\item The $MWN_c$ sends the available accounting 
      parameters $R_{i}^{c}$ to $AS_c$ 

\item $AS_c$ configures its accounting model using 
      the $R_{i}^{c}$ details. $R_{i}^{c}$ consists 
      of $CI_i$ and $TT_i$.

\item  $AS_c$ obtains metered data from $MS_c$ based 
       on the $CI_i$ , computes $SR_{i}^{c}$ using 
       its accounting model and sends it the $MWN_c$.

\item  $MWN_c$ sends $SR_{i}^{c}$, $SR_{i}^{p}$ to the 
       $Comparator_c$, who compares them and returns 
       the decision $decn_i$ to the $MWN_c$. 
\end{enumerate}

If $decni=No$ the $MWN_c$ starts negotiation 
with $RAS_p$ aiming at solving the dispute. 
When $MWN_c$ receives a negotiation response it resets the 
value of $R_{i}^{c}$ according to the evidence 
received from the negotiation response and executes  
steps 2 to 5. When the negotiation is completed 
or $MWN_c$ obtains $decni=Yes$ from the $Comparator_c$, the 
$MWN_c$ sends the $decn_i$ to NR middleware and waits for a 
new an accounting record from the consumer NR middleware. 

\subsection{P2P Protocol for Online Dispute Resolution}
\label{Protocol}
The base of our dispute resolution protocol is the
exchange of SP, EP of the interval under question and
the average TT. The main idea of this protocol that the provider computes the 
resource consumption for each consumption interval 
$CI_i$ and sends it through its NR middleware to the consumer. 
The consumer sends the record to its $RAS_c$ which compares 
it with its own record and in case of dispute it 
starts a negotiation with the $RAS_p$. The $RAS_c$ sends a 
negotiation request to $RAS_p$. The negotiation request 
should contain the accounting parameters of the $RAS_c$. 
The $RAS_p$ compares the consumer parameters with its 
parameters, finds the parameter or parameters that cause 
the conflict and sends them to the $RAS_c$ with a negotiation 
response. The $RAS_c$ uses the provider's parameters to 
compute storage consumption. It then compares the two 
records: if they match it sends a $decni$ 
to the consumer NR middleware and waits for a new 
accounting record. Otherwise, it sends a new negotiation 
request until the end of the protocol. The negotiation 
protocol is completed either when the $RAS_c$ has received 
a stop negotiation message from the $RAS_p$ or the $RAS_c$ 
obtains an agreed decision.

We now show the psudocode of the protocol for
dispute resolution. We use the following notation:

\# $nc_{i}^{c}$= consumer negotiator

\# $decn_i$= decision

\# $SR_{i}^{p}$= provider's storage consumption record containing storage consumption (SC)

\# $SR_{i}^{c}$= consumer's storage consumption record containing storage consumption (SC)

\# TT= transmission time

\# SP= start point, EP= end point

\# $R_{i}^{c}$= consumer's accounting parameters $SP_{i}^{c}$, $EP_{i}^{c}$, $TT_{i}^{c}$

\# $R_{i}^{p}$= provider's accounting parameters $SP_{i}^{p}$, $EP_{i}^{p}$, $TT_{i}^{p}$

\# $NReq_i$= negotiation request $R_{i}^{c}$, $nc_{i}^{c}$

\# $NRes_i$= negotiation response $R_{i}^{p}$, $nc_{i}^{p}$
\begin{figure}[h!]
	\centering
		\includegraphics[width=0.75\textwidth]{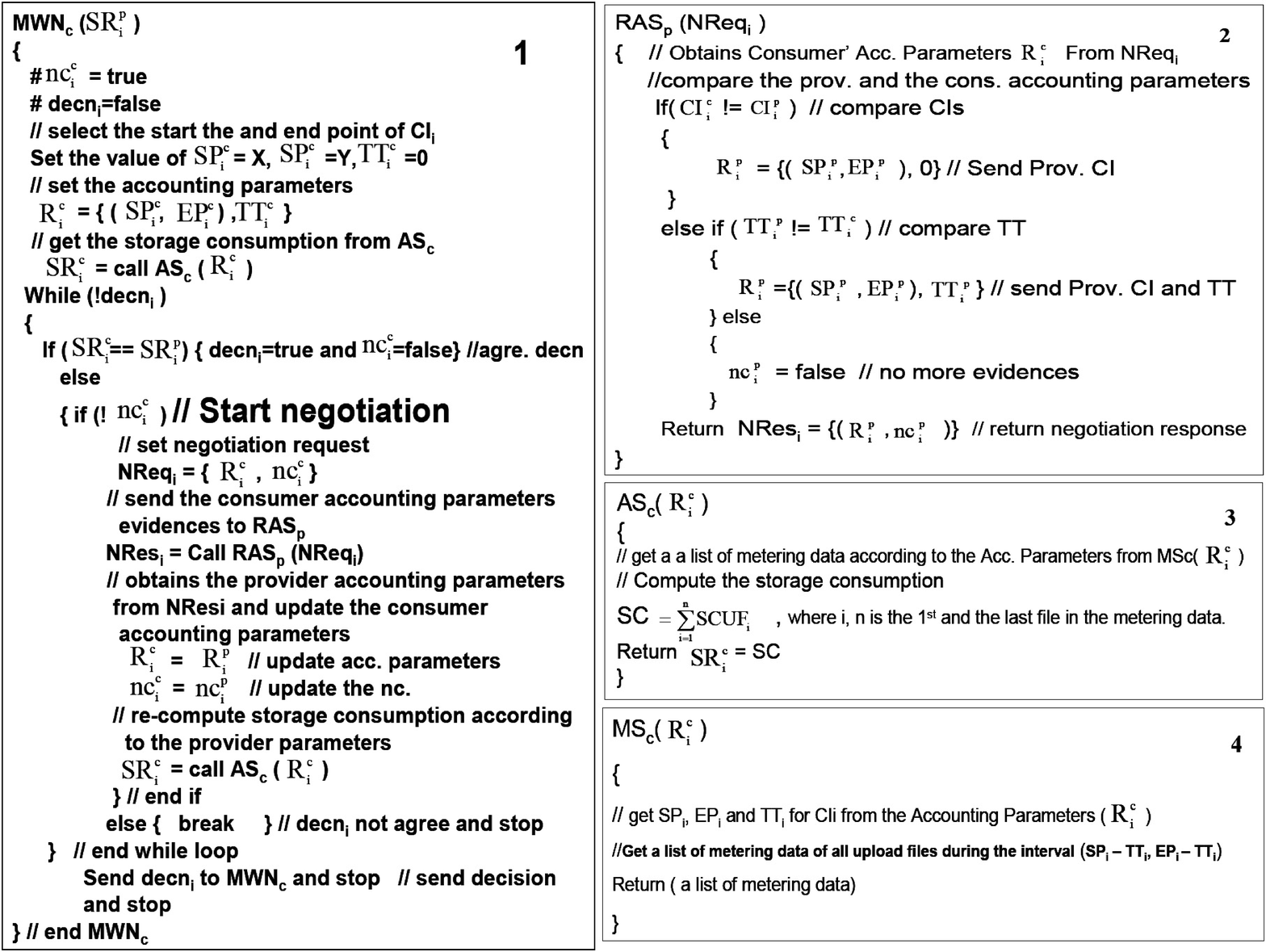}
	\label{fig:figprotocol}
\end{figure}

\section{Related Work}
The idea of bilateral measurement of resource consumption
with online dispute resolution of potential
conflicts was firstly suggested in~\cite{MolinaESBE2008}, 
however, no protocol to solve the problem was discussed;
in this respect, our work can be regarded as a step
forward in this research direction. 
Online Dispute Resolution (ODR) systems based on Utility
Theory have been studied by some authors (see for
example~\cite{ZCXu2008}). A particularity of these ODR systems 
is that they attempt to reach interest-based 
voluntary settlement agreements based on parties'
preferences and tradeoff. In contrast, in our work
disputes are solved---if possible---on the basis
of evidences presented by the two parties involved in
the conflict. Another particularity of these ODR is
that they normally rely on a third party (e.g.~arbitration)
to help solve the dispute; we depart from this idea and
suggest a peer--to--peer conflict resolution protocol. 

The use of middleware interceptors to monitor 
Service Level Agreements--regulated interactions
between two parties is discussed in~\cite{Morgan2005}. 

The problem of deciding the physical location of the 
components of metering services to measure
resource consumption  is related to the concept of monitorability 
discussed in~\cite{Skene2007}. In accordance with these 
authors, a given service level parameter (for example, 
response time, storage consumption, etc.) is 
unmonitorable, monitorable by a trusted third party, 
monitorable by one party, monitorable by both parties. 
In~\cite{MolinaESBE2008} the authors explain that the monitorability 
of a given parameter depends on several factors; among 
the most important ones are the accuracy of the 
metering, the physical location of the application 
(one or many) that affects the parameter, the physical 
location of the metering service and the trust 
assumptions about the metering service and its 
location. 

In~\cite{Alsakran2006} the authors discuss a
protocol to provide  
non-repudiable evidence of services consumption in 
mobile internet services.  
Non-repudible evidence is used by the provider to 
prove the correctness of his bill and allows the 
consumer to verify the service consumption. The
protocol suffers from several limitations. For
example, it involves an online trusted third party to insure fairness of 
the protocol. Likewise, it works for 
time--base accounting only.  Furthermore, the non--repudiable
service increases the number of 
messages exchanged through the network. Consequently, 
the additional messages reduce the effective bandwidth 
for users' data traffic. Moreover, if disputes  
over a service interval appear, the service
is simply terminated; when this happens, the provider 
looses money because the consumer would have already consumed 
part of the service interval. Our work constrasts with this
approach in that we try to solve conflict when they
apper rather than terminating the service.

\section{Conclusion and Future Work}
\label{FutureWork}
In bilateral accounting of resource consumption  
the consumer and the 
provider independently measure resource consumption, 
compare their outcomes and try to agree on a 
a single outcome. In this paper we have discussed when, why 
and where conflicts might happen in bilateral 
accounting for storage consumption. We propose
a peer--to--peer online protocol to be executed
between the consumer and provider to solve
conflics over the consumer's consumption.
In future, we are planning to study other parameters 
(different techniques to collect data about
resource consumption) that might cause disputes 
over storage consumption. 

\bibliographystyle{eptcs}
\bibliography{bibliography}
\end{document}